\documentclass[mnsc,nonblindrev]{informs3} 

\OneAndAHalfSpacedXI 

\usepackage{natbib}
 \bibpunct[, ]{(}{)}{,}{a}{}{,}%
 \def\bibfont{\small}%

	\usepackage{amsmath}
	\usepackage{graphicx}
	\usepackage{enumerate}
	\usepackage{xcolor}
	\usepackage{natbib} 
	\usepackage{url} 
	\usepackage{multirow}
	\usepackage{color}
        \usepackage{soul}
        \usepackage{parallel}
        \usepackage{multicol}
        \usepackage{rotfloat} 
        \usepackage{booktabs}
        \usepackage{caption}
        \usepackage{subcaption}
        \usepackage{lscape}
	\usepackage{amssymb}
	\usepackage{multirow}

\TheoremsNumberedThrough     
\ECRepeatTheorems

\EquationsNumberedThrough    

\MANUSCRIPTNO{}

\renewcommand{\theARTICLETOP}{}

\begin{document}



\RUNTITLE{Priority Thresholds for Equitable Housing Access}

\TITLE{Leveraging Priority Thresholds to Improve Equitable Housing Access for Unhoused-at-Risk Youth}




\ARTICLEAUTHORS{%
\AUTHOR{Yaren Bilge Kaya }
\AFF{Industrial Engineering Department, Northeastern University, Boston, USA  \EMAIL{kaya.y@northeastern.edu}} 
\AUTHOR{and Kayse Lee Maass}
\AFF{Industrial Engineering Department, Northeastern University, Boston, USA  \EMAIL{k.maass@northeastern.edu}}
} 

\ABSTRACT{%
Approximately 4.2 million youth and young adults experience homelessness each year in the United States and lack of basic necessities puts this population at high-risk of being trafficked or exploited. Although all runaway and homeless youth (RHY) are at risk of being victims of human trafficking, certain racial, ethnic, and gender groups are disproportionately affected. Motivated by these facts, our goal is to improve equitable access to housing resources for at-risk RHY in New York City (NYC) by expanding the current housing capacity, while utilizing priority thresholds that guide decisions regarding which youth should start receiving service based on the number of beds idle in the system. Our approach involves an $M/M/N/\{K_j\}+M$ queuing model with many statistically identical servers (beds) and RHY from different demographic groups with limited patience arriving to the a large crisis and emergency shelter in NYC. The queuing model allows us to: (i) investigate the populations and demographics that are facing access barriers, (ii) project the minimum number of beds required to provide a certain global service quality level to all youth, regardless of demographic characteristics, and (iii) use priority thresholds while matching RHY with beds to promote equity. The recommendations regarding the capacity expansion and priority thresholds improves equitable access to this crisis and emergency shelter by decreasing the average number of RHY abandoning the system by 92\%, with a  particular reduction in the abandonment of RHY who are at high-risk of experiencing trafficking.

}%



\maketitle

%


\setlength{\footskip}{24pt}
\pagestyle{plain}

\section{Introduction} \label{sec:intro}

Youth homelessness is a significant and increasing problem around the world \citep{Stephenson-2001, Embleton-2016, Anderson-2016}. In the United States (US), each year approximately 4.2 million runaway and homeless youth and young adults (RHY) between the ages of 14-24 experience homelessness; and, on any given night, approximately 41,000 unaccompanied RHY are unhoused \citep{CFPY}. Approximately half of this population sleep outside, in a car, or some place not meant for human habitation \citep{EndHomelessness-2021}.

Many factors may increase youth's possibility of experiencing homelessness, including the systemic barriers and biases that some youth  face due to their demographic characteristics. Studies show that most minority groups in the US experience homelessness at higher rates than white people \citep{HUD-2013, EndHomelessness-2020}. The most striking disparity can be found among African Americans, who represent 13\% of the general US population but account for 39\% of people experiencing homelessness \citep{Census-2020}. Another study evaluating return to homelessness within less than one year of receiving housing in 16 US communities shows that individuals identifying as Native American or Alaska Native are at significantly higher risk of returning to homelessness relative to those identifying as white; and that identifying as LGBTQ+ is positively correlated with returning to homelessness \citep{Petry-2021}. Research shows that these disparities can be explained by long-standing historical and structural racism, and discrimination against LGBTQ+ communities \citep{Olivet-2019, Romero-2020, Fowle-2022}. 

Being homeless is complex. Even though the consequences of homelessness differ for every youth, research shows that it impacts most youth's physical and mental health negatively. RHY are at great risk of experiencing malnutrition, sexually transmitted diseases, illnesses, injuries, physical abuse, and even homicide \citep{klein-2000}. In addition to possible physical impacts of homelessness, studies show that RHY are more likely to demonstrate high rates of mental health problems such as behavioral problems, depression, anxiety, and self-harm, compared with their housed peers \citep{Cumella-1998, Aviles-2006}. Furthermore, disrupted education is also a common issue for this population, and often makes it particularly challenging to find sustainable employment opportunities \citep{Aratani-2009}. Lack of access to these basic necessities (shelter, affordable and nutritious food, appropriate clothing, healthcare, and more), and experiencing emotional or physical neglect or abuse, coupled with lack of community support structures put RHY in a vulnerable position \citep{Middleton-2018, Hogan-2020, Wright-2021}. 

Traffickers are known to prey upon these vulnerabilities and use coercion, force, or fraud to obtain some type of labor or commercial sex act from youth \citep{Davy-2015}. While doing so, traffickers often provide housing and income opportunities to make it more difficult for youth to leave the trafficking situation \citep{Bigelsen-2013}. Accurate estimates on the prevalence of human trafficking (HT) among homeless youth are difficult to obtain due to the illicit nature of the crime, lack of a standard method of reporting trafficking cases, lack of recognition by authorities, and the tendency of survivors to under-report \citep{Greenbaum-2017}. However, a large 10-city study found that nearly one in five interviewed homeless youth were victims of some type of HT activity \citep{Murphy-2016}. 

Although all RHY are at high-risk of being victims of HT, studies show that certain racial, ethnic, and gender groups are disproportionately affected \citep{Chong-2014}. In the US, domestic child sex trafficking victims are significantly more likely to be a racial/ethnic minority, identify themselves as LGBTQ+, or come from lower socioeconomic class families \citep{Dank-2017, Wolfe-2018, Fedina-2019, Martin-2020}. Another study that took place between 2009-2015 in Florida showed that 49.5\% of the youth who had experienced trafficking were African-American, while only 36.8\% of youth were White and non-Hispanic \citep{Reid-2019}. Furthermore, studies show that demographic characteristics also impact one's risk for labor trafficking victimization \citep{Barrick-2014}. Thus, minorities are both more likely to be homeless and more likely to experience HT than their non-minority peers. 

Root causes of homelessness can be addressed through a range of essential support services, including mental and substance use disorder treatment, employment, and education. These services can serve as measures to prevent initial homelessness, help RHY transition back to stable living situations, reduce youth's vulnerability to trafficking, and help them be part of a healthy community. To make this transition smoother for RHY, there are housing and shelter programs that provide short and long-term temporary residences to RHY. However, only a few of these shelters are able to provide HT trauma-informed care to potential victims and survivors. Furthermore, in most US communities, the number of RHY seeking housing and support services exceeds the number of available resources \citep{Clawson-2009}. Lack of resources in the RHY shelter system often causes long wait times and therefore, delayed care or queue abandonment (youth leaving the system without starting to receive care) \citep{Gibbs-2014}. In this study, we emphasize the significance of increasing emergency, short, and long-term housing resources tailored for potential HT victims and survivors so that youth are not unhoused and vulnerable as a result of long wait lists or capacity limitations.

We seek to reduce disparities in HT rates among homeless youth in NYC by alleviating current capacity limitations and using priority thresholds to decide who can start receiving service based on the number of idle servers in the system. We use a mixed-methods multi-step approach that involves: (i) primary and secondary data collection and analyses to identify the populations that experience disparities in accessing existing scarce housing resources and are at high-risk of experiencing HT, and (ii) a queuing model that incorporates queue abandonment ($M/M/N/\{K_j\}+M$) to further identify the optimal bed capacity decisions and equitable access. The queuing model allows us to project the capacity that needs to be deployed to service providers to adequately meet the needs of RHY, while also identifying priority thresholds to improve equitable access to housing and trauma-informed support services for youth who are at the highest risk of trafficking. We assume that RHY from various vulnerability groups (e.g., multi-class heterogeneous customers) arrive to a large crisis and emergency shelter (e.g., a multi-server queue with many identical servers) to receive housing and support services. To the best of our knowledge, no queuing models with heterogeneous customers currently exist that focus on improving equitable access to public services by determining the optimal capacity expansion while limiting queue abandonment and utilizing priority thresholds.

The remainder of the paper is structured as follows. We provide a review of the literature in Section 2; explain our data collection process in Section 3; propose our queuing model in Section 4; present our simulation model, computational setup and ensuing results and insights regarding applicability in Section 5. We conclude by summarizing our contributions, limitations and future research directions in Section 6.

\section{Literature Review} \label{sec:lit_rev2}
Our work borrows from and contributes to three primary
areas: (i) queue abandonment, (ii) modeling heterogeneous customer behaviour, and (iii) equity and fairness in queuing systems, which have gained increasing attention in recent years. In this section, we discuss their current use, relevance to our study, and limitations. Table \ref{t:queue_summary} summarizes important articles that we explored during this study and helps us point out the literature gaps we address. 

\begin{table}[]
\centering
\caption{Literature relevant to fairness and equity in heterogeneous queuing models with abandonment.}
\begin{tabular}{|l|c|c|c|c|c|}
\hline
\multicolumn{1}{|c|}{\multirow{2}{*}{Study}} & \multirow{2}{*}{\begin{tabular}[c]{@{}c@{}}Queue\\  Abandonment\end{tabular}} & \multicolumn{2}{c|}{Fairness or Equity} & \multicolumn{2}{c|}{Heterogenous} \\ \cline{3-6} 
\multicolumn{1}{|c|}{} &  & \multicolumn{1}{c|}{Customers} & Servers & \multicolumn{1}{c|}{Customers} & Servers \\ \hline
\cite{Alexander-2012} &  & \multicolumn{1}{c|}{} & $\checkmark$ & \multicolumn{1}{c|}{} &  \\  \hline
\cite{Armony-2011} & $\checkmark$ & \multicolumn{1}{c|}{} &  & \multicolumn{1}{c|}{} & $\checkmark$ \\ \hline
\cite{Armony-2010} &  & \multicolumn{1}{l|}{} & $\checkmark$ & \multicolumn{1}{c|}{} & $\checkmark$ \\ \hline
\cite{Armony-2015} & $\checkmark$ & \multicolumn{1}{l|}{} &  & \multicolumn{1}{c|}{} &  \\ \hline
\cite{Avi-itzhak-2004} &  & \multicolumn{1}{c|}{$\checkmark$} & $\checkmark$ & \multicolumn{1}{c|}{} &  \\ \hline
\cite{Azizi-2018} &  & \multicolumn{1}{c|}{$\checkmark$} &  & \multicolumn{1}{c|}{$\checkmark$} &  \\ \hline
\cite{Baccelli-1984} & $\checkmark$ & \multicolumn{1}{l|}{} &  & \multicolumn{1}{c|}{} &  \\ \hline
\cite{Batt-2015} & $\checkmark$ & \multicolumn{1}{l|}{} &  & \multicolumn{1}{c|}{} &  \\ \hline
\cite{Bitran-2008} & $\checkmark$ & \multicolumn{1}{l|}{} &  & \multicolumn{1}{c|}{} &  \\ \hline
\cite{Brandt-1999} & $\checkmark$ & \multicolumn{1}{l|}{} &  & \multicolumn{1}{c|}{} &  \\ \hline
\cite{Brown-2005} & $\checkmark$ & \multicolumn{1}{l|}{} &  & \multicolumn{1}{c|}{} &  \\ \hline
\cite{Dai-2011} & $\checkmark$ & \multicolumn{1}{l|}{} &  & \multicolumn{1}{c|}{} &  \\ \hline
\cite{Gans-2003} & $\checkmark$ & \multicolumn{1}{l|}{} &  & \multicolumn{1}{c|}{} &  \\ \hline
\cite{Garnett-2002} & $\checkmark$ & \multicolumn{1}{l|}{} &  & \multicolumn{1}{c|}{} &  \\ \hline
\cite{Geers-2020} & $\checkmark$ & \multicolumn{1}{l|}{} &  & \multicolumn{1}{c|}{} &  \\ \hline
\cite{Gurvich-2008} &  & \multicolumn{1}{l|}{} &  & \multicolumn{1}{c|}{$\checkmark$} &  \\ \hline
\cite{Gurvich-2009} & $\checkmark$ & \multicolumn{1}{l|}{} &  & \multicolumn{1}{c|}{} & $\checkmark$ \\ \hline
\cite{Hashimoto-2012} &  & \multicolumn{1}{c|}{$\checkmark$} &  & \multicolumn{1}{c|}{} &  \\ \hline
\cite{Larson-1987} & $\checkmark$ & \multicolumn{1}{c|}{} &  & \multicolumn{1}{c|}{} &  \\ \hline
\cite{Leclerc-1995} & $\checkmark$ & \multicolumn{1}{l|}{} &  & \multicolumn{1}{c|}{} &  \\ \hline
\cite{Mandelbaum-2007} & $\checkmark$ & \multicolumn{1}{l|}{} &  & \multicolumn{1}{c|}{} &  \\ \hline
\cite{Mandelbaum-2009} & $\checkmark$ & \multicolumn{1}{c|}{} &  & \multicolumn{1}{c|}{} &  \\ \hline
\cite{Mandelbaum-2012} &  & \multicolumn{1}{c|}{} & $\checkmark$ & \multicolumn{1}{c|}{} & $\checkmark$ \\ \hline
\cite{Palmer-2020} & $\checkmark$ & \multicolumn{1}{c|}{} &  & \multicolumn{1}{c|}{$\checkmark$} & $\checkmark$ \\ \hline
\cite{Rahmattalabi-2022} &  & \multicolumn{1}{c|}{$\checkmark$} &  & \multicolumn{1}{c|}{$\checkmark$} &  \\ \hline
\cite{Shimkin-2004} & $\checkmark$ & \multicolumn{1}{c|}{} &  & \multicolumn{1}{c|}{$\checkmark$} &  \\ \hline
\cite{Su-2006} &  & \multicolumn{1}{c|}{$\checkmark$} &  & \multicolumn{1}{c|}{$\checkmark$} &  \\ \hline
\cite{Suck-1997} & $\checkmark$ & \multicolumn{1}{c|}{} &  & \multicolumn{1}{c|}{} &  \\ \hline
\cite{Tschaikowski-2017} & $\checkmark$ & \multicolumn{1}{c|}{} &  & \multicolumn{1}{c|}{} &  \\ \hline
\cite{Tseytlin-2007} & $\checkmark$ & \multicolumn{1}{c|}{} & $\checkmark$ & \multicolumn{1}{c|}{} & $\checkmark$ \\ \hline
\cite{Whitt-2004} & $\checkmark$ & \multicolumn{1}{c|}{} &  & \multicolumn{1}{c|}{} &  \\ \hline
\cite{Yang-2013} &  & \multicolumn{1}{c|}{$\checkmark$} &  & \multicolumn{1}{c|}{$\checkmark$} &  \\ \hline
\textbf{This article} & $\checkmark$ & \multicolumn{1}{c|}{$\checkmark$} &  & \multicolumn{1}{c|}{$\checkmark$} &  \\ \hline
\end{tabular}
\label{t:queue_summary}
\end{table}

\subsection{Queue Abandonment} \label{s:lit_abandonment}

Queuing theory has been studied rigorously over the last century \citep{Hillier-1967, Halfin-1981, Shortle-2018}. However, queue abandonment is one aspect of human behavior that is poorly understood in queuing theory \citep{Palm-1957, Tschaikowski-2017}. Prior literature has explored the reasons behind abandonment and showed that it's highly affected by long wait times \citep{Leclerc-1995}. Customers who are subjected to long wait times tend to experience feelings such as stress, anxiety, ambiguity, and a sense of wasting time \citep{Suck-1997}. A natural consequence of such feelings is that some customers lose patience and abandon the queue before receiving service. The abandonment probability or the long-run fraction of customers who abandon the system is an important performance measure for most revenue-generating service systems since it directly effects the total revenue \citep{Dai-2011}. In most service settings, abandonment is undesirable and reflects customers' dissatisfaction. Therefore, understanding drivers of queue abandonment is an essential first step in designing better service encounters \citep{Bitran-2008}.

The traditional queuing theory approach to model queue abandonment is the \textit{Erlang-A} model, where each customer has a maximum amount of time they are willing to wait \citep{Garnett-2002, Brown-2005, Mandelbaum-2007}. In the Erlang-A model, the customer either remains in the queue until the service is provided to them or abandons the queue when their actual wait time reaches the maximum amount of time they were willing to wait. This model can also be shown as $M/M/N+M$ using the Kendall notation, where customers arrive according to a  Poisson process, service times follow an exponential distribution, and customers' maximum wait times (e.g., patience) are assumed to be independent and identically distributed (i.i.d.) and follow an exponential distribution \citep{Baccelli-1984, Brandt-1999}. Generally, the choice of modeling abandonment with the Erlang-A approach is driven by tractability concerns \citep{Gans-2003, Whitt-2004}.

Using the steady state approximations, \cite{Mandelbaum-2007, Mandelbaum-2009} present the optimal staffing levels for $M/M/N+M$ queues considering three different operating regimes: efficiency-driven (ED), quality-driven (QD), and quality and efficiency-driven (QED). The authors recommend using the ED regime formulas for non-revenue-generating services that often provide a low-to-moderate quality of service; using the QD regime when the price of missing a customer is significantly high; and using the QED regime to combine high levels of efficiency and service quality. In our study, we evaluate the staffing levels using these three operating regimes through the perspectives of various stakeholders: RHY, governmental decision makers (e.g., the NYC Mayor's Office), and non-profit decision makers (e.g., homelessness assistance program directors), further explained in Section \ref{ss:staffing}. 

\subsection{Heterogeneous Behaviour}\label{s:lit_hetero}
Considering the heterogeneity of customers and servers is a necessary component to model real world conditions. One of the first studies to combine queue abandonment and heterogeneity considers a heterogeneous utility function for customers, where customers act rationally to maximize a utility function that weights service utility against expected waiting cost \citep{Shimkin-2004}. In a different study, \cite{Gurvich-2008} focus on a large-scale service system with multiple customer classes and many statistically identical servers and propose a staffing approach, as well as an approach to match staff with customers while minimizing staffing cost, subject to class-level quality-of-service constraints. Although the authors do not include queue abandonment in their study, they mention the opportunity for extensions to consider it. Therefore, in this study we expand the heterogeneous customer model proposed in \cite{Gurvich-2008} by including queue abandonment, exploring staffing options subject to different class-level quality-of-service constraints, and incorporating equity measures. 

Considering server heterogeneity, \cite{Gurvich-2009} present a family of routing rules called Queue-and-Idleness-Ratio (QIR) rules for large-scale service systems with multiple customer classes and multiple agent pools, each with many agents. 
\citet{Armony-2011} study large-scale service systems with homogeneous impatient customers and heterogeneous servers, and they propose that staffing and routing rules are jointly asymptotically optimal in the heavy-traffic many-server QED and ED regimes, respectively. In a more recent study, \cite{Palmer-2020} present a novel perspective for evaluating the performance of multi-class queuing networks through a combination of operational performance and service quality, and extend the current fluid–diffusion approximation methods to include multiple classes, class transitions and dynamic server allocations. These studies have done a significant job by introducing customer and server heterogeneity to the queue abandonment literature. Nonetheless, to the best of our knowledge, our study is to first to include queue abandonment and heterogeneous customer behaviour considering customer's demographic characteristics. 

\subsection{Equity, Social Justice and Queuing Theory} \label{s:lit_equity}

Gender, racial, and ethnic equity in analytical research means applying tools and practices needed to recognize the LGBQTIA+ communities' and people of color's experiences with unequal power differentials and access to resources and opportunities, while considering historical and current lived realities, including structural racism. 
While equity and fairness measures have been incorporated into some queuing models, more research is needed to broaden their applicability. \cite{Avi-itzhak-2004} propose possible fairness measures, such as variance of the wait times, that enable us to quantitatively measure and compare the level of fairness associated with $G/D/1$ queues. In a more current study, \cite{Alexander-2012} focuse on priority queues in different service settings to evaluate social justice and equity. Focusing on call centers, \cite{Armony-2010} formulate a queuing model to minimize steady-state expected customer wait time subject to a fairness constraint on the workload division. However, none of these earlier studies include impatient customers, nor do they focus on homelessness assistance systems. 

Another important consideration when incorporating equity into queuing models is whether equity among customers or among servers is being assessed. A limited number of studies choose to evaluate the situation from the service providers' perspective. For example,  \cite{Tseytlin-2007} compare different routing policies for patients arriving to a large hospital to find a fair strategy for the medical staff. In this study, the authors include impatient customers served by heterogeneous servers, and they present QED regime approximations for the randomized most idle routing policy. \cite{Mandelbaum-2012} introduce a fair routing strategy for the medical staff by allocating patients to hospital wards while considering the heterogeneous server speeds. The authors aim to find a fair routing algorithm that will ensure a balanced patient allocation by looking at the idleness ratios of servers and the flux ratios between different wards under the homogeneous arrival assumption. In both studies, the authors acknowledge that the customer homogeneity assumption is a limitation of their study. 

Considering equity and fairness from the customer's perspective in healthcare settings,  \cite{Su-2006} present a kidney allocation model considering heterogeneous kidneys with different transplant queues and two alternative social welfare functions: aggregate utility (emphasizing efficiency) and minimum utility across all candidates (emphasizing equity). In another study, \citet{Yang-2013} introduce a queuing simulation model to explore the relationship between intensive care unit bed availability and operating room schedules to maximize the use of critical care resources and minimize case cancellation while providing equity to patients and surgeons. In both of these studies, the definition of equity differs. \cite{Yang-2013} define equity to ensure that the system does not repeatedly reject the same type of surgery, while \cite{Su-2006} guarantee that all kidney transplant candidates have equal chances of receiving an organ. \cite{Su-2006} acknowledge the well-established correlation between race and transplant outcomes. However, while identifying equity measures, neither of these studies include demographic characteristics of patients, and are not able to investigate the outcomes of the policies they recommend. Although not explicitly focused on improving racial and ethnic equity while accessing public services, these studies are instrumental in establishing topics such as fairness and equity within the queuing literature. 

When we look at queuing models that focus on improving equitable and fair access to scarce housing resources, we see two great examples that incorporate racial and ethnic equity. \cite{Rahmattalabi-2022} present fair and interpretable policies to effectively match heterogeneous individuals to scarce housing resources of different types by using causal inference and mixed-integer optimization. In this study, the authors match homeless individuals with eligible housing resources based on the individual's vulnerability score and race and show that their approach improves the rate of exit from homelessness for underserved or vulnerable groups (7\% higher for the Black individuals and 15\% higher for those below 17 years old). \cite{Azizi-2018} propose a flexible data-driven mixed-integer optimization formulation to prioritize heterogeneous homeless youth on a waiting list for scarce housing resources of different types. The authors show that their approach outperforms the current policy and reduces racial discrimination. Although both studies improve equitable access to housing resources, they do not consider the limited patience of this population, and do not address the need to increase the number of available resources. 

We believe that these aforementioned studies were instrumental in establishing queue abandonment, heterogeneity of servers and customers, and fairness in queuing systems as a research field,  although none of the studies consider all of these components at the same time as we do in this study. Thus, this is the first study to our knowledge that addresses gender, racial and ethnic equity by investigating queue abandonment and heterogeneous server and customer behaviour in public service systems.  

\section{Data and Community Partners} \label{sec:data2}
To understand youth homelessness more thoroughly and incorporate a community-based approach to our research, we work closely with community partners who are familiar with the New York City RHY shelter system. As such, the population we focus on in this study is runaway and homeless youth and young adults ages 16-24 in NYC. We have regular meetings with stakeholders, including the NYC Mayor's Office, NYC Coalition for Homeless Youth, and shelter service providers. To learn about RHY's demographic characteristics, work/education background, and vulnerability to HT, as well as their needs, we gather publicly available data from multiple sources \citep{NYC-gov-2021}. Based on these data sources, we create \textit{youth profiles} to represent the larger RHY population in NYC (see \textit{[citation blinded for peer review]}  for more details).  

The homelessness assistance programs in NYC provide different types of beds to RHY who are in need, divided into three main categories: crisis and emergency housing, transitional and independent living, and long-term housing. Traditionally, the crisis and emergency beds are the main entry point to the RHY shelter system and are available to youth for short amounts of time; RHY often can stay in these types of beds up to three months. Additionally, due to the nature of these programs, the support services included with crisis and emergency beds are not as comprehensive as other bed types. They provide basic needs such as accommodation, food, personal, and cleaning supplies, and basic mental and physical health support. In this study, we focus on the entry process of RHY who are at high-risk of experiencing HT to the RHY shelter system. Therefore, we focus only on a large crisis and emergency shelter that has trained staff to provide HT trauma-informed care to RHY in NYC. 

To learn about the existing resources, capacities, demographic groups accepted, types of services offered, and the process flows in shelters, we held meetings with homelessness assistance program stakeholders, explained more in detail in \textit{[citation blinded for peer review]}. In total, to best of our knowledge, there are four major homelessness assistance programs that provide crisis and emergency housing and support services to RHY in NYC, with approximately 265 beds in total. The large shelter we focus on has 164 of all the crisis and emergency beds in NYC, and has trained staff who can provide HT trauma-informed care to HT survivors, victims, and potential victims. 

\subsection{Vulnerability Groups}

We evaluate RHY's vulnerability to HT based on the information we gathered from publicly available data sources, our meetings with key stakeholders, and though our literature review. We reviewed 16 recently published articles (3 literature reviews, 4 reports, and 9 peer-reviewed articles that have been published between years 2010-2022) that focus on youth's vulnerability to HT. The findings from our literature review indicate that previously experiencing exploitative situations, being involved in the child welfare or juvenile justice system, having substance abuse or mental health problems, experiencing emotional or physical abuse or neglect, coming from a lower socioeconomic class or experiencing systemic biases due to being a US gender, sexual orientation, racial, or ethnic minority put RHY at higher risk of experiencing HT. We summarize the risk factors mentioned in each of these studies in Table \ref{t:risk_factors} of Appendix \ref{s:Ap_vulnerabilty}.

Aligned with our findings from our literature review, the meetings with key stakeholders revealed that previously experiencing HT indicates that youth is particularly at high-risk to be re-victimized and should receive rehabilitative services as fast as possible; thus, following their recommendations, we include ``previously experiencing human trafficking" as our most significant risk factor. Although experiencing emotional or physical neglect or abuse is a common risk factor (included in 13 out of 16 articles), studies show that approximately 90\% of RHY experience these concerns \citep{Wolfe-2018}. Therefore, due to their ubiquitous nature in our study population, neglect and abuse are not included as risk factors in our study that differentiate the population from each other. Additionally, meetings with stakeholders revealed that the socioeconomic status of the family is likely to have less effect on RHY during their time spent away from family, on the streets, or in the RHY shelter system; thus, this risk factor is also not included. 

In line with the literature and stakeholder expert opinions regarding youth risk factors for increased risk of HT, our model assumes (i) currently or previously experiencing HT, (ii) having substance abuse or mental health problems, (iii) experiencing systemic biases due to LGBTQ+ status, (iv) involvement in child welfare or juvenile justice systems, and (v) experiencing systemic biases due to belonging to a minority racial or ethnic group in the US are the most important demographic characteristics that affect youth's vulnerability to trafficking (listed as most impactful to least impactful). These vulnerability groups considered in our study are summarized in Table \ref{t:vuln_groups} alongside the approximate proportion the vulnerability group accounts for in the RHY population. These proportions were estimated using various publicly available and primary data sources \citep{Kral-1997, Kipke-1997, Park-2005, Kempf-2007}. Details regarding how we estimate these proportions are given in Tables \ref{t:proportions_vul} and \ref{t:estimates_vul} of Appendix \ref{s:Ap_vulnerabilty}. We validate these assumptions with multiple stakeholders, including service providers, trafficking survivors, and former RHY themselves as defining the vulnerability groups and their proportions requires expertise and careful attention.

\begin{table}[]
\centering
\small
\caption{The vulnerability groups created based on information gathered from primary and secondary data sources, presented with proportion of RHY in NYC that falls under the category.}
\begin{tabular}{p{13cm} c c c}
\toprule
\textbf{Condition} & \textbf{Group}  &  \textbf{\%} \\ \midrule
If youth have previously experienced HT  & A  & 20.00\%\\
If youth have substance abuse or mental health problems but don't meet any of the above conditions  & B   & 22.40\%\\
If youth are LGBTQ+ but don't meet any of the above conditions & C   & 17.28\%\\
If youth are previously or currently involved in child welfare or juvenile justice systems but don't meet any of the above conditions & D   & 8.06\%\\
If youth are a racial or ethnic US minority but don't meet any of the above conditions & E &  24.20\%\\ 
If youth don't meet any of the above conditions & F & 8.06\% \\
\bottomrule
\label{t:vuln_groups}
\end{tabular}
\end{table}

\section{Model Framework}
In this section we explain the different queuing models we consider to project the number of beds required to serve RHY at a large crisis and emergency shelter in NYC and introduce the priority thresholds to prioritize youth who are more vulnerable to trafficking on the shelter wait-lists. Our priority thresholds approach lets youth into a large crisis and emergency shelter based on the number of idle beds. Suppose youth from group A have priority over youth from group B when there are fewer than $K$ beds open in the system. Then, if a youth from group B arrives to the system when there are fewer than $K$ beds open, the group B youth is unable to receive services from that specific shelter until there are greater than $K$ beds open. In our case, we model priority thresholds for a crisis and emergency shelter that has staff with dedicated trauma-informed training specific to HT. Thus, if group A refers to youth who have been trafficked (or are at high-risk) and group B refers to non-trafficked youth, even though youth from Group B would need to wait to get services from this shelter, the youth from Group B could still receive services from other shelters in the area that do not specialize in caring for trafficking survivors. Thus, a priority threshold approach may allow more equitable access for youth who have been trafficked or are at a heightened risk. 

\begin{figure}[h] 
\begin{center}
\includegraphics[height=2in]{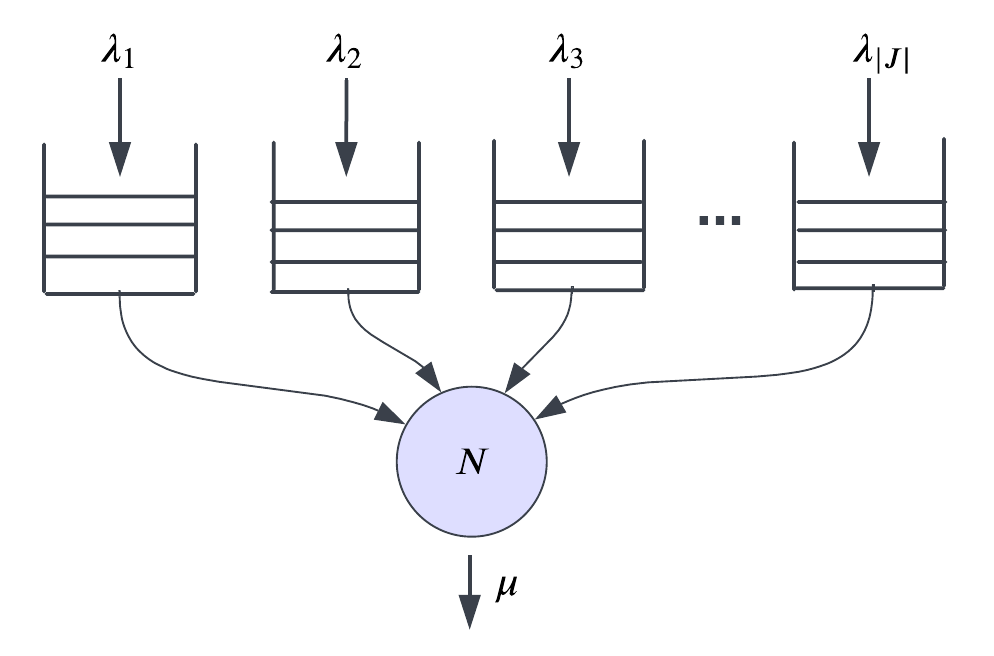}
\caption{The V-Model.} 
\label{f:v-model}
\end{center}
\end{figure}

Using priority thresholds based on number of idle servers in the system is not a new concept; as explained in Section \ref{s:lit_hetero}, 
\cite{Gurvich-2008} previously use this approach to provide varying service quality to heterogeneous customers that are arriving to a call center. In this study, we expand their work by including queue abandonment, equity, and stakeholder perspectives to improve equitable access to housing resources for RHY who are highly vulnerable to trafficking. We model a certain large crisis and emergency shelter system in NYC as a V queuing model by assuming that there are youth arriving from multiple vulnerability groups (multiple customer classes) with limited patience to receive service from many statistically identical servers. This V model is depicted in Figure \ref{f:v-model} \citep{Gross-2008}. 
\vspace{-6mm}

\subsection{Model Formulation}
The sets, parameters, and notation used in our models are included in Table \ref{t:para2} and further detailed in what follows. Let $N$ be the number of beds required at the crisis and emergency shelter we are modeling to serve RHY, and $J$ be the set of vulnerability groups (explained in detail in Section \ref{sec:data2}). We assume that RHY from vulnerability group $j \in J$ arrive to the system, seeking a bed according to a Poisson arrival process with rate $\lambda_j$; thus the aggregate system arrival rate is $\lambda=\sum_{j\in J}\lambda_j$. The service time represents RHY's length of stay at the crisis and emergency shelter and it follows an exponential distribution due to its high variance. The service rate is assumed to be i.i.d. and is represented as $\mu$, while the patience of youth also follows an i.i.d. exponential distribution with abandonment rate $\theta$. The probability of abandonment is denoted $P\{Ab\}$ and the expected wait time in the queue (e.g., the wait that occurs before youth starts receiving service) is $E[W]$. The priority thresholds for vulnerability groups $j \in J$ are represented as $K_j$. 

\begin{table}[]
\centering
\caption{The parameters and notation used in the stochastic program.}
\begin{tabular}{ll}
\toprule
Symbol & Definition \\
\midrule
$J$ & Set of different vulnerability groups $j$, where $j \in J$ \\
$N$ & The number of statistically identical beds in the system  \\
$\lambda_j$ & Arrival rate of vulnerability group $j$, where $j \in J$ \\
$\lambda$ & Aggregate arrival rate of the system, where $\lambda = \sum_{j\in J}\lambda_j$ \\
$\mu$ & Average service rate of the system \\
$\theta$ & Abandonment rate of RHY population \\
$W$ & Steady-state wait time of the entire population \\
$W_j$ & Steady-state wait time of vulnerability group $j$, where $j \in J$ \\
$P\{Ab\}$ & Probability of abandonment of the entire population \\
$P_j\{Ab\}$ & Probability of abandonment of vulnerability group $j$, where $j \in J$ \\
$E[W]$ & Expected wait time of the entire population \\
$K_j$ & Entry threshold for vulnerability group $j$, where $j \in J$ \\
\bottomrule
\end{tabular}
\label{t:para2}
\end{table}

The approach we propose to improve equitable access for at-risk RHY can be summarized in two main steps: (i) estimate the minimum number of crisis and emergency beds ($N$) required at the shelter to provide a certain aggregate level of quality-of-service (QOS) to all RHY, and (ii) estimate the priority thresholds ($K_j$) required in the system to provide equitable access to RHY who are highly vulnerable to HT. Therefore, we propose two different $M/M/N/\{K_j\}+M$ queuing models in \eqref{eq:aban_model}-\eqref{eq:wait_model}, where each model includes a global QOS constraint, as well as an individual QOS constraint set for different vulnerability groups. The global QOS constraints target the whole RHY population and ensure that RHY collectively receive a certain level of service quality from the shelter; while the individual QOS constraints aim to provide a higher quality service than the global service quality for RHY who are more at-risk of trafficking than others. We consider two types of global QOS constraints: limiting the proportion of youth abandonment to be at most $\alpha$ with \eqref{eq:aban_model_1} and ensuring the expected wait time is at most $M$ with \eqref{eq:wait_model_1}. Similarly, two types of individual QOS constraint sets are considered: ensuring that no more than $x_j$ percent of RHY from vulnerability group $j\in J$ wait longer than $T_j$ with \eqref{eq:aban_model_2} and limiting the proportion of RHY from vulnerability group $j \in J$ abandoning the system to be at most $\alpha_j$ with \eqref{eq:wait_model_2}.
\newpage
\begin{multicols}{2}
\begin{subequations}{\label{eq:aban_model}}
\begin{alignat}{3}
& Minimize  \quad     && N                    \\
& Subject \ to \quad  && P\{Ab\} \leq \alpha \label{eq:aban_model_1}\\ 
&                     && P\{W_j  \geq  T_j\} \leq x_j  \quad j= 1,...,|J|-1 \label{eq:aban_model_2}\\
                      &&& N \geq 0 
\end{alignat}
\end{subequations}
\begin{subequations} {\label{eq:wait_model}}
\begin{alignat}{3}
& Minimize \quad && N\\
& Subject \ to \quad  && E[W] \leq M \label{eq:wait_model_1}\\
&&& P_j\{Ab\} \leq \alpha_j \quad j= 1,..,|J|-1 \label{eq:wait_model_2}\\ 
&&& N \geq 0
\end{alignat}
\end{subequations}
\end{multicols}




Calculating the optimal number of beds and priority thresholds is challenging due to the stochastic nature of the system. However, \cite{Gurvich-2008} propose that the queuing performance measures and the optimal staffing levels can be approximated for the $M/M/N/\{K_j\}$ model in a way that simplifies the solution tremendously. The authors provide proofs regarding the similarity between the $M/M/N/\{K_j\}$ approximations and the first come first serve $M/M/N$ (Erlang C model: without queue abandonment). They also mention the possibility of extending these approximations for $M/M/N/\{K_j\}+M$ queues using the $M/M/N+M$ models (Erlang A model: with queue abandonment). Therefore, in this study we focus on the $M/M/N/\{K_j\}+M$ model with queue abandonment, and incorporate the priority thresholds using \cite{Gurvich-2008}'s recommendations. The approximations for the $M/M/N/\{K_j\}+M$ model presented in \cite{Gurvich-2008} are given below in \eqref{eq:approximations}. 
\begin{equation}
    \begin{aligned} \label{eq:approximations}
        P\{W^{M/M/N/\{K_j\}+M}_{|J|}>0\} \approx P\{W_{\lambda, \mu, \theta}^{M/M/N+M}>0\}\\
        E[W^{M/M/N/\{K_j\}+M}] \approx E[W_{\lambda, \mu, \theta}^{M/M/N+M}] \\
        P\{Ab^{M/M/N/\{K_j\}+M}\} \approx P\{Ab_{\lambda, \mu, \theta}^{M/M/N+M}\}
    \end{aligned}
\end{equation}

Based on these approximations, $N$ can be found by using the $M/M/N+M$ formulae while only considering the global quality constraint, and the priority thresholds ($K_j$) can be estimated using the \textit{``Idle-Server-Based Threshold-Priority Policy"} proposed in \cite{Gurvich-2008}. Therefore, in Section \ref{ss:staffing} we explain how to staff $M/M/N/\{K_j\}+M$ queues considering different operating regimes, and in Section \ref{ss:thresholds} we explain how to match RHY with beds while considering the individual QOS constraints for each vulnerability group. 

\subsection{Staffing \boldmath{$M/M/N/\{K_j\}+M$} Queues}{\label{ss:staffing}}
As we explained in Section \ref{sec:lit_rev2}, \cite{Mandelbaum-2007}, \cite{Mandelbaum-2009}, and \cite{Whitt-2004} summarize three different operating regimes to staff $M/M/N+M$ and $M/M/N+G$ queues in their studies: QD, ED, and QED. Quality-driven (QD) regime ensures that participants' wait is negligibly short; thus, the service quality is high. Efficiency-driven (ED) regime ensures that the utilization of the servers is high; therefore, the participants are subjected to long waits and are likely to abandon the queue. In the quality and efficiency driven (QED) regime, there is a delicate combination of high efficiency with a probability of delay that is strictly between zero and one. With the help of these operational regimes, we aim to incorporate three stakeholders' perspectives on the service quality: RHY, non-profit, and the governmental decision makers. Our meetings with stakeholders revealed that RHY's preference aligns best with the QD regime, while homelessness assistance program directors and the NYC Mayor's Office's preferences align more with QED and ED regimes, respectively. 

The staffing formulas for different operating regimes from \cite{Mandelbaum-2007, Whitt-2004} and \cite{Mandelbaum-2009} are given in Equations \eqref{qd-staffing}--\eqref{qed-staffing}, where $N^*_{QD}$, $N^*_{QED}$, and $N^*_{ED}$ represent the number of beds required within the crisis and emergency shelter to provide youth a certain level of service quality, considering the quality-driven, quality and efficiency-driven, and efficiency-driven regimes, respectively. In \eqref{qd-staffing} and \eqref{ed-staffing}, $\gamma$ is equivalent to the probability of abandonment and $R$ is the traffic intensity $R= \lambda/\mu$. The $\beta^*$ parameter given in \eqref{qed-staffing} represents the service quality. More detail on how to obtain $\beta^*$ is provided in the Appendix Section \ref{s:Ap_QED_staffing}.
\begin{equation}\label{qd-staffing}
    N^*_{QD} = R \cdot (1+ \gamma) 
\end{equation}
\begin{equation}\label{ed-staffing}
    N^*_{ED} = R \cdot (1- \gamma) 
\end{equation}
\begin{equation}\label{qed-staffing}
    N^*_{QED} = R + \beta^* \sqrt{R}
\end{equation}

\subsection{Estimating the Priority Thresholds} {\label{ss:thresholds}}


Once the optimal number of beds are determined through the staffing process in Section \ref{ss:staffing}, the priority thresholds can be determined to ensure equitable access to RHY who are at particularly high-risk of experiencing HT. To estimate the priority thresholds ($K_j$) considering the $P\{W_j  \geq  T_j\} \leq x_j$ constraints in \eqref{eq:aban_model}, we use the formulation below in \eqref{eq:itp}, where, $\sigma_j = \sum_{k=1}^{j}\rho_k$, $\rho_j = \sum_{k=1}^{j}{\lambda_k/N\mu}$, and $x \vee y = max\{x,y\}$ \citep{Gurvich-2008}. 

\begin{equation}\label{eq:itp}
    \begin{aligned}
    K_{j+1}-K_j = \left[ \frac{ln(\alpha_jT_j /[P\{W_{j+1}>0\}\hat{\omega}(N^*,\sigma_{j+1}, \sigma_j)])}{ln(\sigma_j)} \right] \vee 0, \text{for } \ j \in \{1, ...,|J|-1\},\\
    \text{where, } \hat{\omega}(N^*, \sigma_j, \sigma_{j-1}) = [N^*\mu(1-\sigma_j)(1-\sigma_{j-1})]^{-1}, \\
    \text{and } P\{W_j >0\}= P\{W_{j+1}>0\}\sigma_j^{K_{j+1}-K_j}, \\
    \text{and } P\{W_{|J|} >0\} = P\{W^{M/M/N+M}_{\lambda, \mu, \theta}(N^*) >0\}.
    \end{aligned}
\end{equation}

To estimate the priority thresholds ($K_j$) considering the $P_j\{Ab\} \leq \alpha_j$ constraint set in \eqref{eq:wait_model}, we expand the formulation given in \cite{Gurvich-2008} by using the $P\{Ab\} = E[W]\theta$ approximation \citep{Mandelbaum-2004, Whitt-2004} and present \eqref{eq:itp2}. The  $\hat{\omega}$ and $P\{W >0\}$ terms presented in \eqref{eq:itp2} are the same as in \eqref{eq:itp}.

\begin{equation}\label{eq:itp2}
    \begin{aligned}
    K_{j+1}-K_j = \left[ \frac{ln(\alpha_jT_j /\theta [P\{W_{j+1}>0\}\hat{\omega}(N^*,\sigma_{j+1}, \sigma_j)])}{ln(\sigma_j)} \right] \vee 0, \text{for } \ j \in \{1, ...,|J|-1\}\\
    \end{aligned}
\end{equation}


\section{Computational Setup and Results} \label{s:computational_setup}

\noindent We demonstrate our model on a large crisis and emergency shelter in NYC using the $M/M/N/\{K_j\}+M$ queuing model with six vulnerability groups explained in Section \ref{sec:data2} and evaluate different global and individual QOS constraint combinations given in \eqref{eq:aban_model} and \eqref{eq:wait_model}. The staffing decisions in \eqref{eq:aban_model} and \eqref{eq:wait_model} depend on the global QOS constraint that limits the proportion of abandonment and the expected wait time, respectively. Our baseline models are shown in \eqref{eq:aban_num} and \eqref{eq:wait_num}. The limits on the \textit{abandonment rates, expected wait time, and the proportion of youth waiting more than a certain amount of time} are set to provide a high quality of service to youth from different vulnerability groups. 

\begin{multicols}{2}
    \begin{equation}\label{eq:aban_num}
        \begin{aligned}
        & Minimize && N \\
        & Subject \ to && P\{Ab\} < 0.04 \\
        &&& P(W_A >1) \leq 0.05 \\
        &&& P(W_B >1) \leq 0.08 \\
        &&& P(W_C >2) \leq 0.05 \\
        &&& P(W_D >2) \leq 0.10 \\
        &&& P(W_E >2) \leq 0.15 \\
        \end{aligned}
    \end{equation}
    
    \begin{equation}\label{eq:wait_num}
        \begin{aligned}
        & Minimize && N \\
        & Subject \ to && E[W] < 1 \\
        &&& P_A\{Ab\} \leq 0.05 \\
        &&& P_B\{Ab\} \leq 0.08 \\
        &&& P_C\{Ab\} \leq 0.10 \\
        &&& P_D\{Ab\} \leq 0.12 \\
        &&& P_E\{Ab\} \leq 0.15 \\
        \end{aligned}
    \end{equation}
\end{multicols}

To estimate the number of beds required in the shelter and the priority thresholds, we gathered information regarding the arrival rates, length of stay of youth, and youth's patience. Data gathered from publicly available sources showed that approximately 1,600 RHY arrive to this crisis and emergency shelter per year and stay on average 60 days \citep{NYC-gov-2021}. We assume that the arrival rate follows a Poisson distribution ($\lambda= 1,600/360 = 4.44 \ youth \ per \ day$) and the service time is assumed to be exponentially distributed ($\mu=0.016 \ youth \ per \ day$). Considering the proportions of the vulnerability groups presented in Table \ref{t:vuln_groups}, the arrival rates of different vulnerability groups correspond to $\lambda_A = 0.89$, $\lambda_B= 1.07$, $\lambda_C = 0.75$, $\lambda_D = 0.24$, $\lambda_E = 0.96$, and $\lambda_F = 0.55$  youth per day. Furthermore, our meetings with stakeholders revealed that the patience of RHY is highly variable yet short; therefore, following our stakeholders' recommendations, we assume that the the maximum amount of time youth are willing to wait follows an exponential distribution with a mean of 2 days ($\theta = 0.5 \ youth \ per \ day$). 

In Table \ref{t:num_beds}, we present the number of beds required in this large shelter to keep the collective proportion of queue abandonment less than 4\%, and to keep the collective expected wait time less than 1 day. The optimal number of beds required is defined as $N^*_{P\{Ab\}<0.04}$ and $N^*_{E[W]<1}$, respectively, using the staffing formulations explained in Section \ref{ss:staffing}. 

\begin{table}[] 
\centering
\caption{The number of beds required to serve RHY, considering different operating regimes and the global quality-of-service constraints. }
\begin{tabular}{lcccc}
\toprule
\textbf{Operating Regime} &  \textbf{ED} & \textbf{QED} & \textbf{QD} & \textbf{Current Capacity} \\ \midrule
\boldmath{$N^*_{P\{Ab\}<0.04}$} & 250 & 270 & 298 & \multirow{2}{*}{164} \\ 
\boldmath{$N^*_{E[W]<1}$} & 135  & 165 & 274 &  \\ \bottomrule
\end{tabular}
\label{t:num_beds}
\end{table}

Currently the crisis and emergency shelter that we focus on has 164 beds in total. Even the most conservative scenario (efficiency driven regime) recommends adding 86 more beds to this shelter to keep the probability of abandonment less than 4\%. Considering the same QOS constraint, the QED regime recommends adding 106 more beds and the QD regime recommends increasing the number of beds in the shelter to 298 beds.  

When the global QOS constraint limits the expected wait time, we see that $N^*_{E[W]<1}$ values are significantly lower than the staffing levels given in $N^*_{P\{Ab\}<0.04}$ due to the limited patience of youth. For example, the $N^*_{E[W]<1}$ model counts a youth who abandons after one day as having a wait time of one day, whereas a youth who waits but is eventually served in two days will have a wait time of two days. Hence, abandoning youth tend to reduce $E[W]$, indicating that a low $E[W]$ may not actually provide adequate access. This is evident in our results; when the system is staffed considering the $N^*_{E[W]<1}$ values, the overall abandonment probability of RHY is equivalent to 50\% for the ED regime, and 23.1\% in QED regime. Our findings show that using the expected wait time global constraint to staff a system with impatient customers is most likely to result in high abandonment rates. Despite the fact that the expected wait time appears low in this model, the system will provide a lower quality of service to youth due to queue abandonment. Thus, while determining the number of beds within the system,  we choose to focus only on Equation \eqref{eq:aban_num} (limiting the collective abandonment probability to be less than 4\%).

After estimating the number of beds required to keep the system abandonment rate below 4\%, we find the entry thresholds ($K_j$) for each vulnerability group using the priority thresholds policy. The least vulnerable group, known as vulnerability Group F, does not have an individual quality of service constraint. However, in addition to the global limit on the abandonment rate, other groups (A-E) have individual service quality constraints limiting the proportion of youth waiting more than a certain amount of time. Given in \eqref{eq:aban_num}, we aim to keep the proportion of RHY from Group A waiting more than a day to less than 5\%; and for Group B, this value goes up to 8\% . When we use the staffing level recommended by the QED regime, as seen in Table \ref{t:thresholds}, youth in  vulnerability groups A - E can receive a bed as soon as a bed opens up; however, a youth from vulnerability Group F needs to wait until at least 25 beds are open since their likelihood to require HT trauma-informed care is lower than their at high-risk peers. Additionally, there are other non-trafficking focused shelters in NYC that can accommodate the needs of RHY from Group F. 

If the staffing decisions are made considering the QD regime, the model eliminates all the entry thresholds since the QD regime staffing level already decreases the wait times for each group significantly. On the other hand, the ED operating regime increases the threshold to 51 beds for youth in vulnerability Group F, shown in Table \ref{t:thresholds}. 

\begin{table}[]
\centering
\caption{The priority thresholds associated with each vulnerability group considering the \boldmath{$N^*_{P\{Ab\}<0.04}$}.}
\begin{tabular}{ccccccc}
\toprule
\textbf{Operating Regime} &  \boldmath{$K_A$} & \boldmath{$K_B$} & \boldmath{$K_C$} & \boldmath{$K_D$} & \boldmath{$K_E$} & \boldmath{$K_F$} \\ 
\midrule
\textbf{ED} & 0 & 0 & 0 & 0 & 0 & 51\\
\textbf{QED} & 0 & 0 & 0 & 0 & 0 & 25\\
\textbf{QD} & 0 & 0 & 0 & 0 & 0 & 0\\
\bottomrule
\end{tabular}
\label{t:thresholds}
\end{table}


\subsection{Baseline Model Results}
We simulate the large crisis and emergency shelter as a $M/M/N/\{K_j\}+M$ queuing model considering the baseline inputs given in Section \ref{s:computational_setup}. For each experiment, we replicate the simulation 100 times by using different youth profiles at each replication to capture the population variability. These profiles are randomly generated using the proportions given in Table \ref{t:estimates_vul} of Appendix \ref{s:Ap_vulnerabilty}. 

In Table \ref{t:comparison_regimes} we present the effect of using different operating regimes: ED, QED, and QD. These regimes help us evaluate different stakeholders' perspectives such as RHY, non-profit, and governmental decision makers considering their desired quality of service. The ED regime results show us that although it is the least costly option, the average proportion of RHY abandonment across 100 replications is higher than other regimes and the system utilization is high. The QD regime results indicate RHY's wait and abandonment is unlikely; however, on average, 23\% of the beds will be idle.  

\begin{table}[]
\centering
\caption{The queuing performance metrics based on the different operating regimes.}
\begin{tabular}{lccc}
\toprule
\textbf{Queuing Performance Metrics} & \textbf{ED} & \textbf{QED} & \textbf{QD} \\
\midrule
System utilization rate & 91.03\% & 84.10\% & 77.40\% \\
Proportion of RHY abandonment & 4.77\% & 2.47\% & 0.00\% \\
Average Wait Time (days) & 0.14 & 0.06 & 0.00 \\
Number of RHY abandoning from at high-risk groups & 23.45 & 0.98 & 0.00 \\
\bottomrule
\end{tabular}
\label{t:comparison_regimes}
\end{table}

Comparison of these operating regimes show us that the QED regime provides a balance between high utilization and high service quality. Therefore, as our baseline model we choose to use the QED regime recommendations where the crisis and emergency shelter has 270 beds and uses a priority threshold of 25 beds for RHY from Group F. In Table \ref{t:baseline_outputs}, we summarize our baseline model results and compare them with two different scenarios: (i) the current system scenario where there are only 164 beds and no priority thresholds, and (ii) the expanded scenario where there are 270 beds in the system and no priority thresholds. 

\begin{table}[]
\caption{Average queuing performance metrics considering 100 replications, based on different scenarios: current system, the scenario with the capacity expansion, and the baseline scenario.}
\begin{tabular}{clccc}
 \toprule
\multicolumn{2}{c}{\begin{tabular}[c]{@{}c@{}}Queuing Performance \\ Metrics\end{tabular}} & \begin{tabular}[c]{@{}c@{}}Current System \\ (164 Beds and \\ No Priority Thresholds)\end{tabular} & \begin{tabular}[c]{@{}c@{}}Only Expanded Model \\ (270 Beds and \\ No Priority Thresholds)\end{tabular} & \begin{tabular}[c]{@{}c@{}}Base Model\\ (270 Beds with \\ Priority Thresholds)\end{tabular} \\
 \midrule
\multicolumn{2}{l}{System utilization   rate} & 99.26\% & 86.21\% & 84.10\% \\
 \midrule
\multirow{7}{*}{\rotatebox[origin=c]{90}{\parbox[c]{2.5cm}{\centering Abandonment (\% and \#)}}}& Whole system & 31.16\% (498 youth) & 2.46\% (39 youth) & 2.47\% (40 youth) \\ \cline{2-5} 
 & Group-A youth & 30.56\% (98 youth) & 2.56\% (8 youth) & 0.06\% (0.18 youth) \\
 & Group-B youth & 31.77\% (122 youth) & 2.63\% (10 youth) & 0.05\% (0.19 youth) \\
 & Group-C youth & 35.00\% (94 youth) & 2.95\% (8 youth) & 0.12\% (0.29 youth) \\
 & Group-D youth & 32.91\% (28 youth) & 1.66\% (2 youth) & 0.01\% (0.01 youth) \\
 & Group-E youth & 28.62\% (99 youth) & 2.08\% (7 youth) & 0.09\% (0.31 youth) \\
 & Group-F youth & 29.11\% (58 youth) & 2.20\% (4 youth) & 19.75\% (39.02 youth) \\
 \midrule
\multirow{7}{*}{\rotatebox[origin=c]{90}{\parbox[c]{2.5cm}{\centering Average Wait (days)}}} & Whole system & 0.624 & 0.054 & 0.055 \\ \cline{2-5} 
 & Group-A youth & 0.507 & 0.040 & 0.006 \\
 & Group-B youth & 0.427 & 0.045 & 0.005 \\
 & Group-C youth & 0.480 & 0.048 & 0.005 \\
 & Group-D youth & 0.503 & 0.037 & 0.006 \\
 & Group-E youth & 0.498 & 0.046 & 0.006 \\
 & Group-F youth & 0.495 & 0.050 & 0.032\\
 \bottomrule
\end{tabular}
\label{t:baseline_outputs}
\end{table}


\begin{figure}
  \centering
    \centering
    \includegraphics[width=1\textwidth]{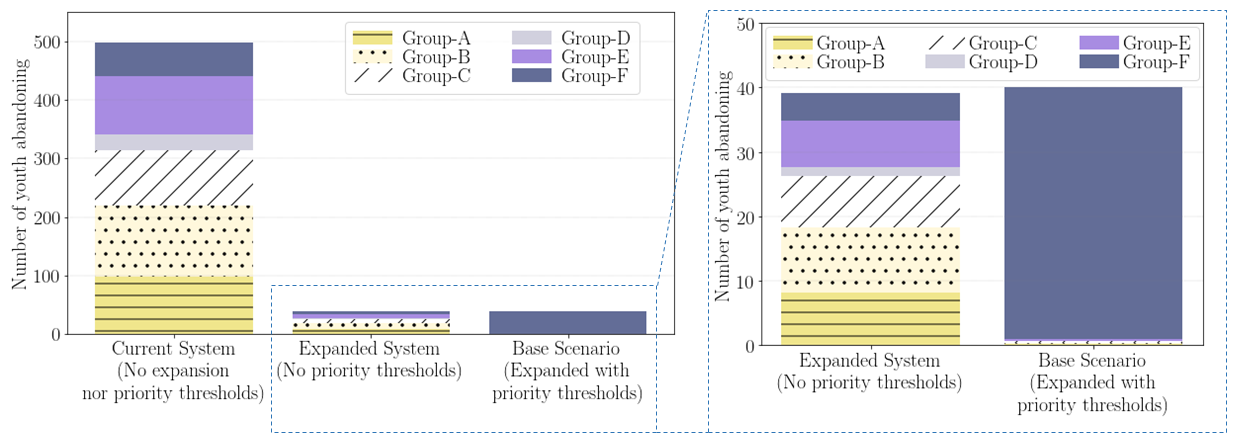}
    \caption{Average number of youth abandoning the queue due to long wait times considering different scenarios, across 100 replications. }
    \label{fig:base_scenarios}
\end{figure}

Our findings (shown in Table \ref{t:baseline_outputs}) indicate that the large NYC shelter is unable to provide timely and appropriate access to housing and support services to RHY with the existing capacity (164 beds). The current system seems to be under-resourced; therefore, the average utilization of beds is close to 100\%. Although the average wait time for a bed is close to half a day for youth from different groups, the average queue abandonment from the system is around 31\% across 100 replications (on average 498 youth leave the queue before receiving a bed due to long-wait times). In this case the proportion of youth abandoning the queue from different vulnerability groups is relatively balanced; the abandonment rates for different groups are between 29\% to 35\% . 

When we increase the number of beds in the system by 106 beds, following the QED regime staffing recommendations, the utilization of the system drops to 86.21\%, and the overall abandonment proportion decreases to 2.46\%. The number of youth abandoning the queue due to long waits decreases from 498 youth to 39, as seen in Figure \ref{fig:base_scenarios}. With the additional 106 beds, the shelter is able to provide housing and support services to 448 more youth with varying vulnerability levels per year. However, in this case, among the 39 youth who abandon the queue, 35 of them qualify as at high-risk of experiencing HT. This model shows that expanding the capacity at this shelter by 106 beds improves access to housing and support services for RHY tremendously, and ensures equal access; but approximately 35 highly-vulnerable youth who arrive to this shelter are still likely to stay unhoused due to lack of access to basic needs. 

To improve equitable access for high-risk youth we propose increasing the number of beds in this shelter by 106 while utilizing priority thresholds for youth from vulnerability Group F, as recommended in Section \ref{t:num_beds}. Introducing the priority thresholds does not change the proportion of queue abandonment (both are approximately 2.465\%). However, it decreases the number of youth abandoning from high-vulnerability groups from 35 to approximately 1 youth, as seen in Figure \ref{fig:base_scenarios}. This model ensures that youth who are at high-risk of experiencing  trafficking have improved access to HT trauma-informed care. While this results in a disproportionate number of youth in vulnerability Group F abandoning this one particular shelter's queue, RHY from vulnerability Group F are the least likely to need to receive the HT specific care at this shelter and can access other shelters in the area that meet their needs.  

\subsection{Sensitivity Analyses}
We now discuss the insights gained through our sensitivity analysis. We vary the model parameters 10 to 20\%, as listed in Table \ref{t:experiments} and evaluate the changes in key queuing performance metrics such as the proportion of queue abandonment, average wait time, and utilization rate. In our sensitivity analyses we consider the QED staffing levels recommended by the model presented in \eqref{eq:aban_num} ($N^*_{P\{Ab\}<0.04}$ =270) and the associated priority thresholds ($K_F = 25$). 

\begin{table}[]
\caption{Experimental parameters varied; bolded values indicate base model parameters.}
\centering
\begin{tabular}{lcc}
\hline
\textbf{Parameter }  & \textbf{Symbol}  & \textbf{Level  }                               \\ 
\hline
Mean Arrival Rate  & {$\lambda$} & 3.55, 4.00, \textbf{4.44}, 4.88, 5.33 youth/day\\
Mean Service Rate & $\mu$ & 0.014, 0.015, \textbf{0.016}, 0.018, 0.021  youth/day\\
Mean Abandonment Rate   & {$\theta$}  &  0.33, \textbf{0.5}, 1 youth/days\\
\hline
\end{tabular}
\label{t:experiments}
\end{table}

\subsubsection{Arrival Rate Sensitivity Analyses}\label{ss:arrival_sens}

Accurate estimates of the demand for housing resources in NYC is challenging to obtain due to hidden nature of this population. Based on the information we gathered from publicly available resources and meetings with stakeholders, we estimate the demand for this large crisis and emergency shelter to be 1600 youth per year ($\lambda=4.44 \ youth/day$). To see the robustness of our capacity expansion and priority threshold recommendations, we perform sensitivity analysis on the arrival rate by increasing and decreasing the baseline scenario arrival rate by 10\% and 20\%. 

\begin{figure}
  \centering
  \begin{subfigure}[t]{0.47\textwidth}
    \centering
    \includegraphics[width=1\textwidth]{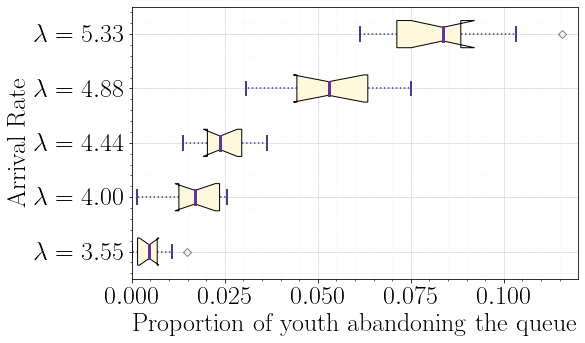}
    \caption{Proportion of youth abandoning the queue.}
    \label{fig:arrival_abandonment}
  \end{subfigure}
  \begin{subfigure}[t]{0.47\textwidth}
    \centering
    \includegraphics[width=1\textwidth]{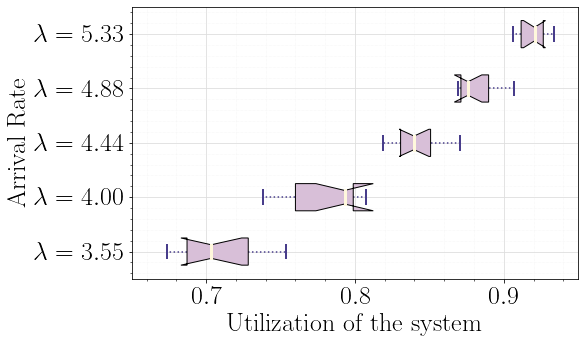}
    \caption{Utilization of the system.}
    \label{fig:arrival_utilization}
  \end{subfigure}  
  \caption{Performance metrics of the system based on the arrival rate change considering 100 replications.}
  \label{fig:arrival}
\end{figure}

Considering the capacity expansion and the priority thresholds recommended in Section \ref{s:computational_setup}, as seen in Figure \ref{fig:arrival}, when the arrival rate increases by 20\% (from 4.44 to 5.33 youth/day), the average proportion of queue abandonment increases to 8\% and the average system utilization increases to 95\%. Although this increase might not be significant in other service systems it is undesirable within service systems that are tailored to provide services to vulnerable populations. The increase in utilization rate is likely to result in faster staff burnout \citep{Baker-2007}, and the increase in queue abandonment may result in increased vulnerabilities. Furthermore, the decrease in arrival rates decreases both the utilization of the system and the queue abandonment, as expected. The decrease in both these performance metrics does not affect the service quality perceived by RHY; however, the 70\% utilization rate indicates that the system is over-resourced. Therefore, the accurate estimates of the arrival rates are important in order to achieve desired service quality levels when implemented. 

\subsubsection{Service Rate Sensitivity Analyses} \label{ss:service_sens}

The number of days a youth stays in a shelter varies depending on the individual youth, housing type, and organization. Due to various reasons, such as feeling limited by organizational restrictions, avoiding conflict and abuse, relapse, health concerns, or finding stable housing, youth may stop receiving housing services \citep{Donley-2021}. Although estimating the length of stay of RHY is challenging, our stakeholder meetings revealed that RHY on average stay at this large crisis and emergency shelter for 60 days. Therefore, we assume that the RHY's service time follows an exponential distribution with the mean of 60 days ($\mu =0.016 
\ youth/day$). To see the effect of service time on queuing performance metrics, we vary the mean service time of youth by 10\% and 20\%. 

\begin{figure}
  \centering
  \begin{subfigure}[t]{0.47\textwidth}
    \centering
    \includegraphics[width=1\textwidth]{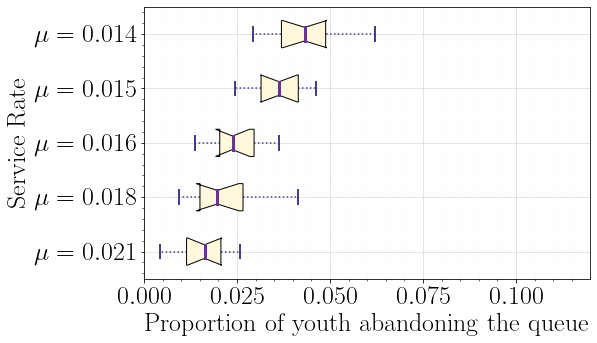}
    \caption{Proportion of youth abandoning the queue. }
    \label{fig:service_abandonment}
  \end{subfigure}
  \begin{subfigure}[t]{0.47\textwidth}
    \centering
    \includegraphics[width=1\textwidth]{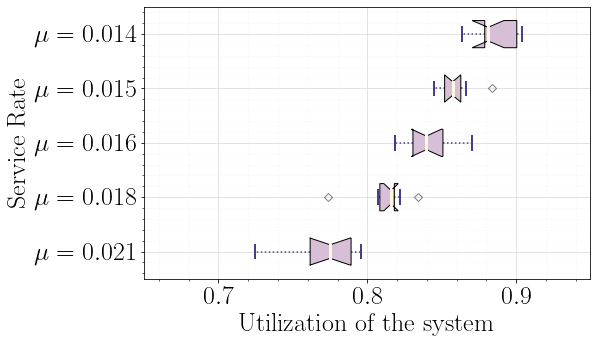}
    \caption{Utilization of the system. }
    \label{fig:service_utilization}
  \end{subfigure}  
  \caption{Performance metrics of the system based on the service rate change considering 100 replications.}
  \label{fig:service}
\end{figure}

The metrics presented in Figure \ref{fig:service} show that the change in service rates does not affect the average abandonment proportion nor the average utilization of the system as much as the arrival rate change. Even if youth's average service time in the system increases from 60 days to 72 days (20\% increase, $\mu = 0.014$), the average abandonment proportion only increases by 2\% and the average utilization of the system does not exceed 90\%. Therefore, our recommendations regarding the capacity expansion and the priority thresholds provide desirable service quality levels for both RHY and managers, even if youth's service time in the system increases or decreases by 10-20\%.  

\subsubsection{Abandonment Rate Sensitivity Analyses}\label{ss:abandonment_sens} 

\begin{figure}
  \centering
  \begin{subfigure}[t]{0.47\textwidth}
    \centering
    \includegraphics[width=1\textwidth]{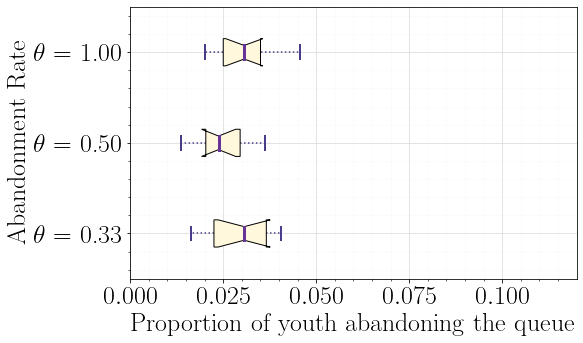}
    \caption{Proportion of youth abandoning the queue. }
    \label{fig:aban_abandonment}
  \end{subfigure}
  \begin{subfigure}[t]{0.47\textwidth}
    \centering
    \includegraphics[width=1\textwidth]{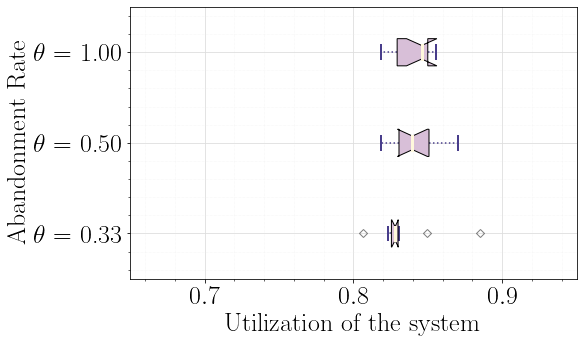}
    \caption{Utilization of the system. }
    \label{fig:aban_utilization}
  \end{subfigure}  
  \caption{Performance metrics of the system based on the abandonment rate change considering 100 replications.}
  \label{fig:abandonment}
\end{figure}

The motivation behind incorporating abandonment within this queuing system is that RHY are known to be highly impatient; yet, how long they are willing to wait to receive housing and support services is variable. Therefore, following our stakeholder recommendations, we assume that RHY's patience follows an exponential distribution with the mean of 2 days. In Figure \ref{fig:abandonment}, we change the abandonment rate of RHY to 1 day and 3 days to see the effect of this change on queuing performance metrics. The change in RHY's patience seems not to impact the average proportion of youth abandoning the system and the average utilization rate as much as changes in the arrival rate or the service rate. Thus, both the average system utilization and the average abandonment proportions are likely to stay in desired service levels, even if RHY's patience is subject to change. 

\begin{figure}
    \centering
    \includegraphics[width=0.5\textwidth]{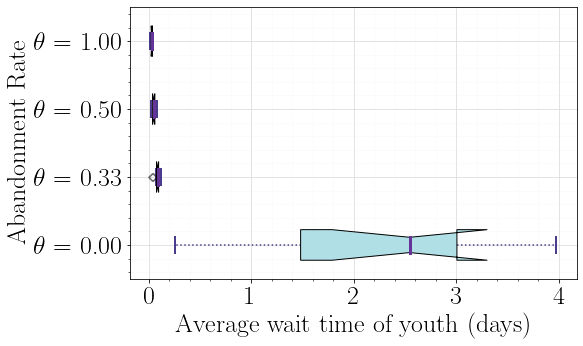}
    \caption{Average wait time of RHY based on varying abandonment rates. }
    \label{fig:abandonment_wait}
\end{figure}

We perform additional analyses on RHY's patience and introduce a scenario where youth have unlimited patience ($\theta = 0.00$). When RHY's patience is unlimited (RHY will stay in the queue until they receive service; they will not lose patience and abandon the queue), we see the most striking change in the average wait times. In Figure \ref{fig:abandonment_wait}, when RHY's patience is unlimited, the average wait time of the system increases from approximately 0 days to 2.5 days. Therefore, to design better service encounters, it is important to know when and where to incorporate queue abandonment into mathematical modeling. The incorrect use of queue abandonment in mathematical modeling may result in under-resourcing and lower service quality than expected.  

\subsection{Mathematical Insights}
In this study, we illustrate the importance of incorporating queue abandonment accurately while modeling service systems by considering a scenario where RHY have unlimited patience and are willing to wait until they receive services, regardless of the wait. This scenario in Section \ref{ss:abandonment_sens}, shows us that if our assumptions regarding RHY's patience were wrong, and RHY are, in fact, patient, the average wait time would be around 2.5 days, which is less than the service quality we desire to achieve. Therefore, the decisions that are based on a queuing model with unnecessary queue abandonment is likely to result in lower quality of service, increased wait times of participants, and increased server utilization due to under-staffing. Furthermore, in the systems where the participants are often impatient, the lack of modeling queue abandonment may result in higher costs due to over-staffing, and lower server utilization. Therefore, understanding the drivers of queue abandonment and modeling it accurately is critical to inform staffing decisions. 

When we focus on systems where the participants are impatient, we see that the selection of performance metrics to decide the optimal staffing levels is important. To illustrate this importance, we compare models presented in \eqref{eq:aban_num} and \eqref{eq:wait_num}, where the staffing decision are based on the proportion of queue abandonment and the expected wait time of the system, respectively. In our case, when we consider the model presented in \eqref{eq:aban_num} to determine the number of beds needed in this large crisis and emergency shelter, we see that the number of required beds is significantly higher than the results of \eqref{eq:wait_num}, presented in Table \ref{t:num_beds} (105 more beds, considering QED regime). We show that limiting the expected wait time to be less than 1 day at this system where RHY are on average likely to wait for 2 days causes 23.1\% of RHY to abandon the system due to limited patience. Therefore, we emphasize the significance of evaluating different queuing performance metrics to decide on the staffing levels, especially in queuing models with impatient participants. 

\subsection{Applicability to Practice: Stakeholder Perspectives}
To understand the current system operations better and evaluate the applicability of our threshold policy approach we conduct structured surveys with former-RHY (3 individuals) and service providers (4 individuals) from various organizations in NYC. Regarding the current system, we ask participants (i) whether they think that shelters in NYC are able to provide equal access to US minorities compared to their non-minority peers, (ii) whether shelters prioritize RHY, and (iii) to compare various HT risk factors that affect RHY. Unsurprisingly, all participants (7/7 individuals) agree that the current system is unable to provide equal access to RHY from US minority groups. Furthermore, service providers indicate that shelters do prioritize youth based on various factors such as likelihood to exit homelessness, immigration status, or HT victimization history. However, one service provider's response emphasizes the lack of standardized protocol to inform service providers regarding how to prioritize RHY on wait-lists. The response given below shows the need for a standardized priority policy within homelessness assistance programs:

\textit{``Well, they're suppose to [prioritize] but often times they do not adhere to protocol due to needing to meet funding requirements of keeping the beds filled. In my opinion there is no direct answer to this question because each provider operates differently but overall, most marginalized individuals do not get placed or services based on their direct ask/needs.''}

In addition to our questions related to current system operations, we also explain the priority thresholds policy that we introduce in this study. To address the racial, gender, and ethnic disparities in the current system, we shift our focus from equality to equity. We ask whether the priority thresholds policy (i) can improve equitable access for at high-risk RHY and (ii) is fair to RHY who are at lower vulnerability groups. Out of 7 respondents, 4 of them indicate that the priority policy can, in fact, improve equitable access to housing resources for at high-risk RHY and that the policy is fair for RHY who are at lower risk vulnerability groups. The remainder of the respondents emphasize the need to expand the housing resources in NYC to serve all RHY, rather than prioritizing youth based on the HT vulnerability due to challenges regarding identifying potential HT victims and survivors. Further comments of the participants highlight the particular need for trauma-informed care in all NYC shelters:

\textit{``Current victims of human trafficking and youth who are at high-risk of human trafficking need specific resources. Most of the shelters available are not specific for youth survivors of human trafficking. At shelters that are trauma-informed, culturally inclusive and LGBTQ+ friendly, it is important to give priority for those who are experiencing violence and abuse. This policy truly highlights the need for specific human trafficking housing resources for youth."}

Considering the feedback we received from our stakeholders and survey respondents, we believe that the priority thresholds policy can improve equitable access for those who are the most vulnerable. However, identifying vulnerability to HT takes careful attention and expertise. Therefore, we acknowledge that this policy change should require a training component to ensure that staff working in shelters can properly identify current and potential victims of HT.

\section{Conclusions}
This study is motivated by the fact that runaway and homeless youth (RHY) who are US gender, sexual orientation, racial, and ethnic minorities are at higher risk of experiencing human trafficking (HT) than their non-minority peers. Although access to housing and support services decreases RHY's vulnerability to HT, the demand, especially from HT survivors, exceeds the capacity of the existing housing resources in most US communities. Therefore, in this study we propose a community-based, data-informed, and interdisciplinary approach to alleviate current capacity limitations by using priority thresholds to improve equitable access to a crisis and emergency shelter in New York City (NYC) for at-risk RHY.  

Our literature review, meetings with stakeholders, and surveys with service providers reveal that prioritizing RHY on wait-lists considering different attributes is common \citep{Rahmattalabi-2022, Azizi-2018, Chan-2018}; however, housing systems often focus on individual's likelihood of exiting homelessness rather than improving equitable access or reducing the individual's vulnerability. Therefore, in this study to ensure \textit{equitable access rather than equal access} to those who are at high-risk of experiencing HT, we use priority thresholds presented in \cite{Gurvich-2008}. This approach helps us guide decisions regarding which RHY should start receiving service based on the number of beds idle in the system. We expand \cite{Gurvich-2008} by incorporating queue abandonment, equity, and stakeholder perspectives. To the best of our knowledge, this is the first queuing study that incorporates priority thresholds, queue abandonment, heterogeneous customer behaviour, and equity simultaneously.

To evaluate RHY's vulnerability to trafficking and the effect of various environmental and societal HT risk factors, we examine the literature focusing on RHY and HT risk factors, hold meetings with stakeholders such as NYC Mayor's Office and Coalition for Homeless Youth, and conduct structured surveys with multiple service providers and former RHY. Our findings indicate that experiencing trafficking, having substance abuse or mental health problems, experiencing systemic biases due to being a US gender, sexual orientation, racial, or ethnic minority, being involved in child welfare and juvenile justice systems put RHY at high-risk to experience trafficking. Therefore, provision of timely access to housing resources is particularly important while dealing with RHY who experienced any of the aforementioned risk factors. 

The results we obtain from our study and the infromation gathered from stakeholder meetings confirm that the existing resources at this large crisis and emergency shelter are not sufficient to provide a high quality of service to RHY. Our results indicate that currently on average 1 out of every 3 RHY abandons the shelter queue due to long wait times, while the average utilization of the system is around 100\%. Therefore, our model recommends increasing the number of beds at this shelter from 164 beds to 270 beds, to limit the proportion of youth abandoning the queue to 4\%. This expansion reduces the number of RHY abandoning the system per year by 92\% (498 to 39 youth/year); however, the proportion of youth abandoning the queue from different vulnerability groups remains similar to the current system; leading 90\% of the queue abandonment to be from the high-risk category (36 out of 39 youth/year). To decrease the number of at high-risk RHY on the streets due to lack of access, we propose restricting the entrance of RHY from the least vulnerable group (RHY who has not experienced the listed risk factors) until there are more than 25 beds idle in the system. With this approach we reduce the number of youth abandoning the queue from at high-risk groups from 90\% to 2.5\%. This major reduction in queue abandonment highlights that our approach, in fact, is an effective way to reduce the number of potential HT victims.

Furthermore, we conduct sensitivity analyses on key model parameters such as the arrival rate, service rate, and abandonment rate of the system to evaluate the robustness of our recommendations. The results show that the accurate estimates of the arrival rate plays a significant role to provide the desired quality of service to RHY. However, the changes in service rate and abandonment rates do not affect the queuing performance metrics as much. 

We are aware that our approach is highly relevant to policy; therefore, we conduct surveys and hold meetings with former RHY (3 individuals) and service providers (4 individuals) to learn about the applicability of our methodology. The survey results show that 100\% of the respondents identify the need to expand capacity for short-to-long term housing resources in NYC. Furthermore, the majority of respondents mention the need to increase the ``specific resources that are tailored for current victims of human trafficking and youth who are at high-risk of experiencing human trafficking". 

While this study specifically focuses on a single crisis and emergency shelter within NYC, there remains great potential for organizations in other locations to implement this methodology. In the future, this methodology can also be used for a group of shelters who provide HT trauma-informed care by assuming that the arrivals to the system and the number of beds within the system are aggregated. However, the use of a coordinated entry system within the city or the municipality is likely to play a big role for the implementation process, due to lack of access to information regarding the number of available beds at each shelter at any given time. Furthermore, this study can also be extended mathematically by considering heterogeneous server behaviour.  

In conclusion, this study provides preliminary evidence of the value of incorporating the needs of vulnerable populations and equity into humanitarian operations research problems. Our approach benefits government and nonprofit decision-makers by offering a means to effectively improve equitable access to scarce housing resources, while readily enabling sensitivity analyses to examine the effect of demand and service duration changes.




\ACKNOWLEDGMENT{The authors would like to thank the New York City Mayor's Office and New York Coalition for Homeless Youth for their insights and all of the former RHY and RHY service providers that helped inform this study.}

\bibliographystyle{informs2014}

\bibliography{references}

\newpage
\setcounter{page}{1}
\begin{APPENDICES}

\section{Estimation of Vulnerability Group Proportions} \label{s:Ap_vulnerabilty}

We estimate the proportion of RHY from different vulnerability groups by using the information gathered through publicly available data sources, literature review, and stakeholder meetings \citep{NYC-gov-2020, NYC-gov-2021}. Table \ref{t:proportions_vul} shows the likelihood of RHY experiencing the individual vulnerability factors with the sources. 

\begin{table}[h!]
\centering
\caption{The estimates of vulnerability group proportions, given with corresponding references.  }
\begin{tabular}{lccl}
\toprule
\textbf{Vulnerability   Categories} & \textbf{Yes} & \textbf{No} & \textbf{References} \\ \midrule
HT   Victim & 20\% & 80\% & \begin{tabular}[c]{@{}l@{}}Stakeholder meeting, \\ \cite{Pilnik-2018}\end{tabular} \\ \midrule
\begin{tabular}[c]{@{}l@{}}Substance Abuse or \\ Mental Health Problems\end{tabular} & 30\% & 70\% & \begin{tabular}[c]{@{}l@{}}Stakeholder meeting, \\ \cite{Kral-1997}, \\ \cite{Kipke-1997}\end{tabular} \\ \midrule
LGBTQ+ & 30\% & 70\% & \cite{Wolfe-2018} \\ \midrule
\begin{tabular}[c]{@{}l@{}}Child   Welfare  or \\ Juvenile Justice Involvement\end{tabular} & 30\% & 70\% & \begin{tabular}[c]{@{}l@{}}\cite{Park-2005}, \\ \cite{Culhane-2007}, \\ \cite{Kempf-2007} \end{tabular} \\ \midrule
US   minority & 55\% & 45\% & Stakeholder meeting \\ \bottomrule
\end{tabular}
\label{t:proportions_vul}
\end{table}

Due to a general lack of consistent, comprehensive data, we assume that these demographic characteristics are independent from each other. Estimates of the proportion of youth that belong to different vulnerability groups is given in Table \ref{t:estimates_vul}. Summation of percentages given in Table \ref{t:estimates_vul} for each vulnerability group is equal to the proportions given in Table \ref{t:vuln_groups}.

\begin{table}[]
\centering
\caption{The estimates of RHY proportions from different vulnerability groups.}
\begin{tabular}{ccccccc}
\toprule
\multicolumn{5}{c}{\textbf{Possible   Combinations}} & \multirow{2}{*}{\textbf{\begin{tabular}[c]{@{}c@{}}Vulnerability \\ Group\end{tabular}}} & \multirow{2}{*}{\textbf{\%}} \\ \cline{1-5}
\textbf{\begin{tabular}[c]{@{}c@{}}HT \\ Victim \end{tabular}} & \textbf{\begin{tabular}[c]{@{}c@{}}Substance \\ Abuse or \\ Mental Health \\ Problems\end{tabular}} & \textbf{LGBTQ+} & \textbf{\begin{tabular}[c]{@{}c@{}}Child Welfare  \\ or Juvenile \\ Justice \\ Involvement\end{tabular}} & \textbf{\begin{tabular}[c]{@{}c@{}}US \\ minority \end{tabular}} &  &  \\ \midrule
0 & 0 & 0 & 0 & 0 & F & 12.348\% \\
0 & 0 & 0 & 1 & 0 & D & 5.292\% \\
0 & 0 & 1 & 0 & 0 & C & 5.292\% \\
0 & 1 & 0 & 0 & 0 & B & 5.292\% \\
0 & 0 & 1 & 1 & 0 & C & 2.268\% \\
0 & 1 & 0 & 1 & 0 & B & 2.268\% \\
0 & 1 & 1 & 0 & 0 & B & 2.268\% \\
0 & 1 & 1 & 1 & 0 & B & 0.972\% \\
0 & 0 & 0 & 0 & 1 & E & 15.092\% \\
0 & 0 & 0 & 1 & 1 & E & 6.468\% \\
0 & 0 & 1 & 0 & 1 & C & 6.468\% \\
0 & 1 & 0 & 0 & 1 & B & 6.468\% \\
0 & 0 & 1 & 1 & 1 & C & 2.772\% \\
0 & 1 & 0 & 1 & 1 & B & 2.772\% \\
0 & 1 & 1 & 0 & 1 & B & 2.772\% \\
0 & 1 & 1 & 1 & 1 & B & 1.188\% \\ \midrule
1 & 0 & 0 & 0 & 0 & A & 3.087\% \\
1 & 0 & 0 & 1 & 0 & A & 1.323\% \\
1 & 0 & 1 & 0 & 0 & A & 1.323\% \\
1 & 1 & 0 & 0 & 0 & A & 1.323\% \\
1 & 0 & 1 & 1 & 0 & A & 0.567\% \\
1 & 1 & 0 & 1 & 0 & A & 0.567\% \\
1 & 1 & 1 & 0 & 0 & A & 0.567\% \\
1 & 1 & 1 & 1 & 0 & A & 0.243\% \\
1 & 0 & 0 & 0 & 1 & A & 3.773\% \\
1 & 0 & 0 & 1 & 1 & A & 1.617\% \\
1 & 0 & 1 & 0 & 1 & A & 1.617\% \\
1 & 1 & 0 & 0 & 1 & A & 1.617\% \\
1 & 0 & 1 & 1 & 1 & A & 0.693\% \\
1 & 1 & 0 & 1 & 1 & A & 0.693\% \\
1 & 1 & 1 & 0 & 1 & A & 0.693\% \\
1 & 1 & 1 & 1 & 1 & A & 0.297\% \\ \midrule
 &  &  &  &  &  \textbf{Total:} & 100\% \\ \bottomrule
\end{tabular}
\label{t:estimates_vul}
\end{table}

\begin{landscape}
\begin{table}[h!]
\centering
\caption{Risk factors that increase vulnerability to human trafficking in youth}
\begin{tabular}{lcccccccc}
\toprule
Study & \begin{tabular}[c]{@{}c@{}}Previously\\Trafficked\end{tabular} & LGBTQ+ & \begin{tabular}[c]{@{}c@{}}Child Welfare \\or Juvenile \\ Justice \\ Involvement\end{tabular} & \begin{tabular}[c]{@{}c@{}}Substance \\ Abuse \\ or Mental \\ Health \\ Problems\end{tabular} & \begin{tabular}[c]{@{}c@{}}Emotional or \\ Physical Abuse \\ or Neglect\end{tabular} & \begin{tabular}[c]{@{}c@{}}US Minority \\ Status\end{tabular} & \begin{tabular}[c]{@{}c@{}}Socioeconomic \\ Status of \\  Family\end{tabular} \\ \midrule
\cite{Gibbs-2014} & $\checkmark$ & $\checkmark$ & $\checkmark$ &  & $\checkmark$ &  & $\checkmark$ \\
\cite{Reid-2019} & $\checkmark$ & $\checkmark$ & $\checkmark$ & $\checkmark$ & $\checkmark$ &  & $\checkmark$ \\
\cite{Dank-2017} & $\checkmark$ & $\checkmark$ &  & $\checkmark$ & $\checkmark$ &  & $\checkmark$ \\
\cite{Gibbs-2018} & $\checkmark$ & $\checkmark$ & $\checkmark$ & $\checkmark$ & $\checkmark$ & $\checkmark$ & $\checkmark$ \\
\cite{Devries-2020} &  &  &  &  & $\checkmark$ &  &  \\
\cite{Jaeckl-2021} &  &  & $\checkmark$ & $\checkmark$ & $\checkmark$ & $\checkmark$ & $\checkmark$ \\
\cite{Olsen-2021} &  & $\checkmark$ &  & $\checkmark$ & $\checkmark$ & $\checkmark$ & $\checkmark$ \\
\cite{Fedina-2019} & $\checkmark$ & $\checkmark$ & $\checkmark$ &  & $\checkmark$ & $\checkmark$ &  \\
\cite{Choi-2015} &  & $\checkmark$ &  & $\checkmark$ &  & $\checkmark$ &  \\
\cite{Wolfe-2018} &  & $\checkmark$ & $\checkmark$ & $\checkmark$ & $\checkmark$ & $\checkmark$ & $\checkmark$ \\
\cite{Twis-2020} &  &  & $\checkmark$ & $\checkmark$ & $\checkmark$ & $\checkmark$ & $\checkmark$ \\
\cite{Macias-2018} &  &  & $\checkmark$ & $\checkmark$ &  &  &  \\
\cite{Greenbaum-2016} &  &  &  & $\checkmark$ & $\checkmark$ &  & $\checkmark$ \\
\cite{Artadi-2010} & $\checkmark$ &  & $\checkmark$ & $\checkmark$ & $\checkmark$ &  & $\checkmark$ \\
\cite{Schwarz-2019} & $\checkmark$ &  &  &  &  & $\checkmark$ & $\checkmark$ \\
\cite{Moore-2020} & $\checkmark$ &  & $\checkmark$ & $\checkmark$ & $\checkmark$ &  &  \\ \midrule
Number of   mentions & 8 & 8 & 10 & 12 & 13 & 8 & 11 \\ \bottomrule
\end{tabular}
\label{t:risk_factors}
\end{table}
\end{landscape}

\newpage
\section{Staffing \boldmath{$M/M/N+M$} Queues Considering QED Regime}\label{s:Ap_QED_staffing}

Let $n_{\lambda}^*$ be the staffing level of the quality and efficiency-driven regime while keeping the probability of abandonment less than a certain level $M$:

\begin{align} {\label{eq:qed-calc1}}
n^*_{\lambda} = argmin_n\{P_{n,\lambda}\{Ab\}\leq M\}
\end{align}

\noindent In Equation \eqref{eq:qed-calc6}, $h_{\phi}(x)$ is the hazard rate of the standard normal distribution; $\Phi(x)$ is the standard normal cumulative distribution function, $\bar{\Phi} = 1-\Phi(x)$ is the survival function, and $\phi(x)$ is the density function.

\noindent Then, for M/M/N+G queues (Erlang-A model with generally distributed customer patience), the abandonment rate is defined as follows in Mandelbaum and Zeltyn (2007) and Mandelbaum and Zeltyn (2009):

\begin{align}{\label{eq:qed-calc2}}
P\{Ab\} \sim \frac{1}{\sqrt{\lambda}} P_a(\beta) P_w(\beta)
\end{align}
\noindent where, 

\begin{align}{\label{eq:qed-calc4}}
    P_a(\beta) = \sqrt{g_0}\cdot(h_{\phi}\hat{(\beta)}-\hat{\beta})
\end{align}

\begin{align}{\label{eq:qed-calc3}}
    P_w(\beta) = \left[1+\sqrt{g_0/\mu}\cdot \frac{h_{\phi}\hat{(\beta)}}{h_{\phi}{(-\beta)}} \right]^{-1}
\end{align}

\begin{align}{\label{eq:qed-calc5}}
    \hat{\beta}= \beta \cdot \sqrt{\mu/g_0}
\end{align}
\begin{align}{\label{eq:qed-calc6}}
    h_{\phi}(x) = \frac{\phi(x)}{1-\Phi(x)} = \frac{\phi(x)}{\bar{\Phi}}
\end{align}

 To obtain exponential patience time distributions instead of general, Zeltyn and Mandelbaum (2005) recommends simply replacing the patience density at the origin $g_0$, with the mean of the exponentially distributed patience $\theta$ in Equations \eqref{eq:qed-calc1}-\eqref{eq:qed-calc6} \citep{Zeltyn-2005}. Therefore, after we combine Equations \eqref{eq:qed-calc1}-\eqref{eq:qed-calc6} and replace $g_0$ with $\theta$, we find the probability of abandonment in Equation \eqref{eq:qed-calc7}. 

\begin{align}{\label{eq:qed-calc7}}
P\{Ab\} \sim \frac{1}{\sqrt{\lambda}} \cdot \sqrt{\theta}\cdot(h_{\phi}(\beta \sqrt{\mu/\theta})-\beta \sqrt{\mu/\theta}) \cdot \left[1+\sqrt{\theta/\mu}\cdot \frac{h_{\phi}(\beta \sqrt{\mu/\theta})}{h_{\phi}{(-\beta)}} \right]^{-1}
\end{align}

\noindent Here, $\beta^*$ represents the service quality level and is the unique solution to:
\begin{align} {\label{eq:qed-calc8}}
M \sqrt{\lambda} = \sqrt{\theta}\cdot(h_{\phi}(\beta \sqrt{\mu/\theta})-\beta \sqrt{\mu/\theta}) \cdot \left[1+\sqrt{\theta/\mu}\cdot \frac{h_{\phi}(\beta \sqrt{\mu/\theta})}{h_{\phi}{(-\beta)}} \right]^{-1}
\end{align}

\noindent Finally, the staffing level recommended by QED regime is given below in \eqref{qed-staffing-Apendix}.
\begin{equation}\label{qed-staffing-Apendix}
    N^*_{QED} = R + \beta^* \sqrt{R}, \ where, \ R= \lambda/\mu
\end{equation}

\section*{Calculating \boldmath{$P\{Ab\}$} and \boldmath{$E[W]$} based on \boldmath{$M/M/N+M$} formulae}

In \cite{Mandelbaum-2007}, $P\{Ab\}$ and $E[W]$ are defined as Equation \eqref{eq:probabiliy_abandonment1}

\begin{align}{\label{eq:probabiliy_abandonment1}}
    P\{Ab| W>0\} = \frac{1}{\rho A (\frac{N\mu}{\theta}, \frac{\lambda}{\theta})}+1-\frac{1}{\rho} \\
    \text{where, } \rho = \frac{\lambda}{N\mu}\\
    A(x,y) = \frac{xe^y}{y^x}\gamma(x,y) \quad and, \quad \gamma(x,y) = \displaystyle \int^{y}_{0} t^{x-1}e^t dt, \quad x>0, y>0
\end{align}

Considering these, $P\{Ab| W>0\}$ is equal to \eqref{eq:probabiliy_abandonment2}

\begingroup
\begin{equation} \label{eq:probabiliy_abandonment2}
    P\{Ab| W>0\} = \frac{1}{\rho \frac{(\frac{n\mu}{\theta}) e^{\lambda/\theta} }{\frac{\lambda}{\theta}^{\frac{n\mu}{\theta}}} \cdot [\sum^{\frac{n\mu}{\theta} -1}_{k=0}{ (-1)^{\frac{n\mu}{\theta}-1-k}\frac{(\frac{n\mu}{\theta}-1)!}{k!}(\lambda/\theta)^k] e^{\lambda/\theta} }  }
\end{equation}
\endgroup

And the $E[W]$ is can be estimated using the $P\{Ab\} = E[W]\theta$ approximation. 

\end{APPENDICES}

\end{document}